\documentclass[twocolumn,amsmath,amssymb,aps,prl,showpacs]{revtex4}
\usepackage{amsmath}
\usepackage{amssymb}
\usepackage{txfonts}
\usepackage{graphicx}
\usepackage{appendix}
\usepackage{epstopdf}
\usepackage{dcolumn}
\usepackage{bm}
\usepackage[colorlinks=true,citecolor=blue,linkcolor=magenta]{hyperref}
\begin{document}

\title{Magnon laser based on Brillouin light scattering
}

\author{Zeng-Xing Liu}\email{liuzx@dgut.edu.cn}
\affiliation{School of Electronic Engineering $\&$ Intelligentization, Dongguan University of Technology, Dongguan, Guangdong 523808, China}
\author{Hao Xiong}\email{haoxiong1217@gmail.com}
\affiliation{School of Physics, Huazhong University of Science and Technology, Wuhan 430074, People's Republic of China}
\date{\today}
\begin{abstract}
An analogous laser action of magnons would be a subject of interest and is crucial for the study of nonlinear magnons spintronics. Here, we demonstrate the magnon laser behavior based on Brillouin light scattering in a ferrimagnetic insulator sphere which supports optical whispering gallery modes and magnon resonances.
We show that the excited magnon plays what has traditionally been the role of the Stokes wave and is coherently amplified during the Brillouin scattering process, making magnon laser possible. Furthermore, the stimulating excited magnon number increasing exponentially with the input light power can be manipulated by adjusting the external magnetic field. In addition to providing insight into magneto-optical interaction, the study of magnon laser action will help to develop novel technologies for handling spin-wave excitations and could affect scientific fields beyond magnonics. Potential applications range from preparing coherent magnon sources to operating on-chip functional magnetic devices.
\end{abstract}

\maketitle

Magnons, the quasiparticle of spin-wave excitation, are well known for its favorable compatibility with a wide range of carriers such as microwaves \cite{photon1,photon2,photon4}, phonons \cite{phonon1,phonon11,phonon2}, and superconducting qubits \cite{qubit1,qubit2}.
Recently, experimental works have reported that the coupling between magnons and microwave photons can reach a strong and even ultrastrong regime \cite{strong1,strong2,strong4}.
In addition, the yttrium iron garnet (YIG) ferrimagnetic material has a high magnetic quality factor and supports magnons with long coherence times, which envisions the promising candidates for quantum information processing \cite{time2}.
In particular, the YIG sphere supports optical whispering gallery modes (WGMs) and magnon resonances \cite{WGM2,WGM,WGM1}.
Simultaneously, the magnon-photon interaction in the YIG sphere gives rise to Brillouin light scattering \cite{BL3}, which is a well-established tool for the characterization of magnonic features and magnetic dynamics.

Brillouin light scattering, in a quantum mechanical language, is essentially an inelastic scattering of light excited by various quasiparticles \cite{waves}, such as phonons, polarons, or magnons within a medium.
In the YIG sphere, akin to 
the mechanical effect of light \cite{OMIT1}, the coherent conversion between the magnons and photons can be achieved via the Brillouin scattering process even though the large frequency mismatch between the optical and magnon modes.
Previous work has shown that the Brillouin light scattering in the YIG sphere will be greatly enhanced
when the triple-resonance condition is satisfied \cite{Brillouin1}, i.e., the input and output optical modes are resonance with the frequency difference given by the magnon mode.
Some coherent effects, ranging from optomagnonically induced transparency  \cite{EIT} and magnon-induced high-order sideband generation \cite{chaos1,chaos2} to magnon blockade effect \cite{blockade} have been revealed.

An analogous laser action of magnons would be a subject of interesting for a broad range of physics from the areas of spintronics, magnonics, and photonics.
The purpose of this work is to establish a theoretical framework for magnon laser action. Some important characteristics of magnon laser, for instance, the gain factor,
threshold power, laser control, and environment temperature have been analyzed.
We found that the excited magnon in the YIG sphere plays what has traditionally been the role of the Stokes wave and is coherently amplified during the Brillouin scattering process, making magnon laser action possible \cite{OMIT1}.
The effective magnon gain is proportional to the input light power, while the threshold power of the magnon laser is inversely proportional to the
square of the magnon-photon coupling strength.
Moreover, we shown that one can achieve the magnon laser control by adjusting the external magnetic field.
In addition, we theoretically evaluated the possibility of observing the magnon laser action at room temperature under the current experimental conditions \cite{WGM,WGM1}.
Beyond their fundamental scientific significance, our results offer attractive prospects for preparing coherent magnon sources \cite{cool1,cool2}, designing magnon-laser amplifiers, and manipulating on-chip magnetic devices \cite{device1,device2}.

\begin{figure}[htbp]
\centering
\includegraphics [width=1\linewidth] {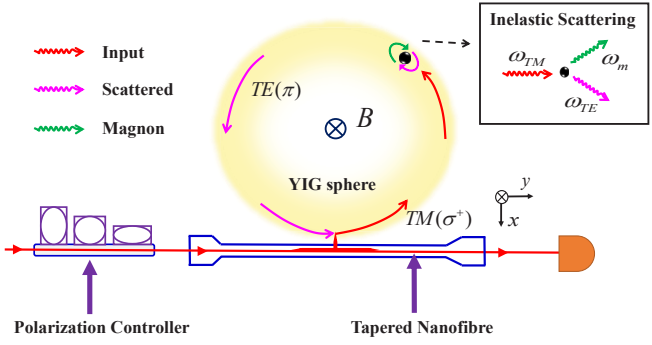}
\caption{Schematic diagram of the physical model.
Similar to the optical WGMs, the YIG ferrimagnetic sphere also supports optical WGMs and magnon resonances.
The inelastic scattering is a process of three-particle interaction, i.e., $\omega_{\rm{M}} \rightarrow \omega_{\rm{E}} + \omega_{\rm{m}}$, where, respectively, the input photon with $\sigma^{+}$ polarization in the TM mode and the scattered photon with $\pi$ polarization in the TE mode.}
\label{fig:1}
\end{figure}

The physical model is schematically shown in Figure \ref{fig:1}, in which a micrometer-scale YIG sphere supports optical WGMs and magnon resonances.
A bias magnetic field B perpendicular to the plane of the WGMs (x-y plane) is applied to saturate the magnetization.
The frequency of the uniform Kittel (magnon) mode in the YIG sphere can be flexibly tuned by adjusting the external magnetic field \cite{Q2}, i.e., $\Omega_{\rm{m}} = \varrho \rm{H_{m}}$, where $\varrho = 2\pi\times28$ $\rm{GHz/T}$ is the gyromagnetic ratio and $\rm{H_{m}}$ is the magnetic field strength.
The input light is introduced through a polarization controller and then evanescently coupled to the optical WGMs via a tapered nanofiber.
The transverse-electric (TE) mode and the transverse-magnetic (TM) mode in the optical WGMs with the same mode index have distinct frequency differences due to the geometrical birefringence, which is demonstrated to be valid regardless of the circulation direction of the photon \cite{birefringence1}.
Assuming that the input light is adjusted to couple to the TM mode (TM input), as shown in Fig. \ref{fig:1}, the light in the resonator is $\sigma^{+}$ polarized \cite{WGM,WGM1}.
The inelastic scattering process occurs in the YIG sphere, i.e., the annihilation of a $\sigma^{+}$-polarized photon in the TM mode accompanied by the creation of a magnon and a down-converted red-sideband photon with $\pi$ polarization in the TE mode.
The interaction between the two light photons and one magnon can be well described by the Hamiltonian
${\rm{H_{int}}} = \hbar g(a_{\rm{M}}a_{\rm{E}}^{\dagger}m^{\dagger} + a_{\rm{M}}^{\dagger}a_{\rm{E}}m)$, where $a_{\rm{M (E)}}$ ($a_{\rm{M (E)}}^{\dagger}$) and $m$ ($m^{\dagger}$) are the annihilation (creation) operators of the photon in the TM (TE) mode and the magnon mode, respectively.
$g = \mathbb{V}\frac{c}{n_{\rm{r}}}\sqrt{\frac{2}{n_{\rm{spin}}\mathbf{V}}}$ is the magnon-photon coupling strength \cite{WGM1}, with the Verdet constant $\mathbb{V}$, the vacuum speed of light $c$, the refractive index $n_{\rm{r}}$, the spin density $n_{\rm{spin}}$, and the YIG sphere volume $\mathbf{V}$.
In this circumstance, the magnon mode plays what has traditionally been the role of the Stokes wave and is coherently amplified during the Brillouin scattering process.

In order to model the magnon laser action more rigorously, we proceed from the Hamiltonian including the TE and TM modes and the magnon mode, as well as the interaction Hamiltonian
\begin{eqnarray}\label{equ:01}
  \rm{H} &=& \hbar\Omega_{\rm{M}}a_{\rm{M}}^{\dagger}a_{\rm{M}}+\hbar\Omega_{\rm{E}}a_{\rm{E}}^{\dagger}a_{\rm{E}}+
  \hbar\Omega_{\rm{m}}{m}^{\dagger}{m}\nonumber \\
  &&+\hbar g(a_{\rm{M}}a_{\rm{E}}^{\dagger}m^{\dagger} + a_{\rm{M}}^{\dagger}a_{\rm{E}}m),
\end{eqnarray}
where $\Omega_{\rm{M (E)}}$ and $\Omega_{\rm{m}}$ are, respectively, the frequency of the TM (TE) mode and the magnon mode.
We assume that the TM mode is driven by a input light with the frequency $\omega_{\rm{M}}$, which allows us to replace the operator $a_{\rm{M}}$ with a classical field $\sqrt{n_{\rm{M}}}e^{-i\omega_{\rm{M}}t}$, where
\begin{eqnarray}
  n_{\rm{M}} &=& \frac{\Upsilon_{\rm{M}}{\rm{P_{in}}}}{\hbar\omega_{\rm{M}}[\Delta_{\rm{M}}^{2}+(\frac{\gamma_{\rm{M}}}{2})^{2}]},
\end{eqnarray}
indicates the average number of photons in the TM mode with the detuning $\Delta_{\rm{M}} = \Omega_{\rm{M}} - \omega_{\rm{M}}$, the input light power $\rm{P_{in}}$ and the decay rate $\gamma_{\rm{M}} = \kappa_{\rm{M}} + \Upsilon_{\rm{M}}$ ($\kappa_{\rm{M}}$ the intrinsic decay rate and $\Upsilon_{\rm{M}}$ the external coupling).
Next we would like to turn to a rotation framework subject to an unitary transformation $\mathbf{U}(t) = \exp( i\omega_{\rm{M}}a_{\rm{M}}^{\dagger}a_{\rm{M}}t+i\omega_{\rm{E}}a_{\rm{E}}^{\dagger}a_{\rm{E}}t
+i\omega_{\rm{m}}{m}^{\dagger}{m}t$) with $\omega_{\rm{M (E)}}$ and $\omega_{\rm{m}}$ the frequency of the input (output) field for the TM (TE) mode and the magnon mode. The Hamiltonian describes the coupling between the TE and the magnon modes driven by the TM mode, therefore, can be obtained as
\begin{eqnarray}\label{equ:02}
  \rm{H} &=& \hbar\Delta_{\rm{E}}a_{\rm{E}}^{\dagger}a_{\rm{E}}+
  \hbar\Delta_{\rm{m}}{m}^{\dagger}{m}\nonumber \\
  &&+\hbar g\sqrt{n_{\rm{M}}}(a_{\rm{E}}^{\dagger}m^{\dagger}e^{-i\delta t} + a_{\rm{E}}m e^{i\delta t}),
\end{eqnarray}
where $\Delta_{\rm{E}(\rm{m})} = \Omega_{\rm{E}(\rm{m})} - \omega_{\rm{E}(\rm{m})}$ denotes the detuning from the TE mode and magnon mode resonances, respectively.
Under the condition of triple resonance, i.e., $\omega_{\rm{M}} - \omega_{\rm{E}} = \omega_{\rm{m}}$, the index factor $e^{\pm i\delta t} \equiv 1$ for $\delta \equiv \omega_{\rm{M}} - \omega_{\rm{E}} - \omega_{\rm{m}} = 0$.
The evolution of the photons and magnons can be well described by the following coupling equations
\begin{eqnarray}\label{equ:1}
\dot{a}_{\rm{E}} &=& (-i\Delta_{\rm{E}}-\frac{\gamma_{\rm{E}}}{2}){a}_{\rm{E}}-ig\sqrt{n_{\rm{M}}}m^{\dagger}
+\sqrt{\kappa_{\rm{E}}}\psi_{\rm{E}}^{(\rm{in})}(t),\nonumber\\
  \dot{m} &=&  (-i\Delta_{\rm{m}}-\frac{\gamma_{\rm{m}}}{2})m-ig\sqrt{n_{\rm{M}}}{a}_{\rm{E}}^{\dagger}+\sqrt{\kappa_{\rm{m}}}\varphi^{(\rm{in})}(t),
\end{eqnarray}
where $\gamma_{\rm{E}}$ ($\kappa_{\rm{E}}$) and $\gamma_{\rm{m}}$ ($\kappa_{\rm{m}}$) are, respectively, the total (intrinsic) decay rates of the TE and magnon modes.
$\psi_{\rm{E}}^{(\rm{in})}(t)$ and $\varphi^{(\rm{in})}(t)$ represent the thermal noise terms of the photon and magnon modes,
and characterized by the following temperature-dependent correlation functions \cite{Noise}  $\langle\psi_{\rm{E}}^{(\rm{in})}(t)\psi_{\rm{E}}^{(\rm{in})\dagger}(t')\rangle$ = $[n_{\rm{th}}(\omega_{\rm{E}})+1]\delta (t-t')$, $\langle\psi_{\rm{E}}^{(\rm{in})\dagger}(t)\psi_{\rm{E}}^{(\rm{in})}(t')\rangle$ = $[n_{\rm{th}}(\omega_{\rm{E}})]\delta (t-t')$, and $\langle\varphi^{(\rm{in})}(t) \varphi^{(\rm{in})\dagger}(t')\rangle$ = $[m_{\rm{th}}(\omega_{\rm{m}})+1]\delta (t-t')$, $\langle\varphi^{(\rm{in})\dagger}(t)\varphi^{(\rm{in})}(t')\rangle$ = $[m_{\rm{th}}(\omega_{\rm{m}})]\delta (t-t')$, where $n_{\rm{th}}(\omega_{\rm{E}}) = [{\rm{exp}}(\frac{\hbar\omega_{\rm{E}}}{\rm{K_{B}T}})-1]^{-1}$, and $m_{\rm{th}}(\omega_{\rm{m}}) = [{\rm{exp}}(\frac{\hbar\omega_{\rm{m}}}{\rm{K_{B}T}})-1]^{-1}$ with the Boltzmann constant $\rm{K_{B}}$ and the ambient temperature $\rm{T}$, are, respectively, the equilibrium means thermal photon and magnon numbers.
For an experiment temperature of T $\sim$ 100 mK \cite{strong4}, the thermal magnon numbers are $m_{\rm{th}} \approx 0.0083$, which is far less than the emitted magnon numbers, and thus the environment thermal noises can be safely ignored.




Transferring the variables to the rotating frame by setting $\tilde{a}_{\rm{E}} = {a}_{\rm{E}} e^{i\Delta_{\rm{E}} t}$ and $\tilde{m} = me^{i\Delta_{\rm{m}} t}$, we shall have
\begin{eqnarray}\label{equ:2}
\dot{\tilde{a}}_{\rm{E}} &=&-\frac{\gamma_{\rm{E}}}{2}\tilde{a}_{\rm{E}}-ig\sqrt{n_{\rm{M}}}e^{i(\Delta_{\rm{E}}+\Delta_{\rm{m}})t}\tilde{m}^{\dagger}, \nonumber\\
\dot{\tilde{m}} &=& -\frac{\gamma_{\rm{m}}}{2}\tilde{m}-ig\sqrt{n_{\rm{M}}}e^{i(\Delta_{\rm{E}}+\Delta_{\rm{m}})t}\tilde{a}_{\rm{E}}^{\dagger}.
\end{eqnarray}
We can adiabatically eliminate the optical mode degrees of freedom since $\gamma_{\rm{m}} \ll \gamma_{\rm{E}}$
\cite{adiabatic}, and then by solving for the antiderivative, we can obtain that
\begin{figure}[htbp]
\centering
\includegraphics [width=0.9\linewidth] {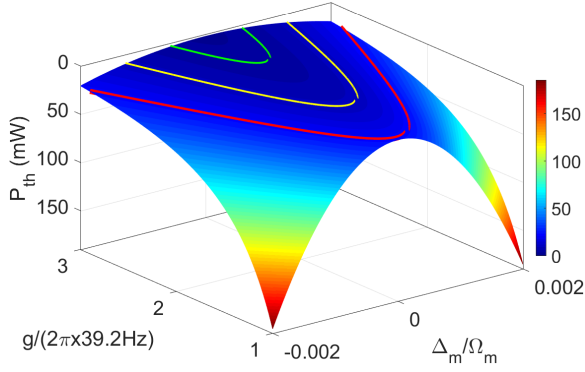}
\caption{The threshold power of the magnon laser $\rm{P_{th}}(\rm{mW})$ as a function of the magnon-photon coupling strength $g/(2\pi\times39.2$ $\rm{Hz})$ and the detuning from the magnon mode $\Delta_{\rm{m}}/\Omega_{\rm{m}}$.
The red, yellow, green solid lines correspond to the threshold power $\rm{P_{th}}$ = 22, 10, 4 $\rm{mW}$, respectively.
The simulation parameters we use are \cite{WGM,WGM1}  $\Omega_{\rm{M}}/2\pi$ $\approx$ $\Omega_{\rm{E}}/2\pi$ = 300 $\rm{THz}$, $\Omega_{\rm{m}}/2\pi$ = 10 $\rm{GHz}$, $\gamma_{\rm{M}}/2\pi$ = $\gamma_{\rm{E}}/2\pi$ = 15 $\rm{MHz}$, $\gamma_{\rm{m}}/2\pi$ = 1 $\rm{MHz}$, $\Upsilon_{\rm{M}} = \gamma_{\rm{M}}/2$, and $\Delta_{\rm{M}} = \Delta_{\rm{E}} = 0$.}
\label{fig:2}
\end{figure}
\begin{eqnarray}\label{equ:3}
  \tilde{a}_{\rm{E}} &=& e^{-\int\frac{\gamma_{\rm{E}}}{2}dt}\int -ig\sqrt{n_{\rm{M}}}\tilde{m}^{\dagger}e^{i(\Delta_{\rm{E}}+\Delta_{\rm{m}})t}e^{\int\frac{\gamma_{\rm{E}}}{2}dt}dt \nonumber\\
&&=\frac{-ig\sqrt{n_{\rm{M}}}}{i(\Delta_{\rm{E}}+\Delta_{\rm{m}})+\frac{\gamma_{\rm{E}}}{2}}e^{i(\Delta_{\rm{E}}+\Delta_{\rm{m}})t}\tilde{m}^{\dagger}.
\end{eqnarray}
Substituting this solution into Eq. (\ref{equ:2}) yields
\begin{eqnarray}\label{equ:4}
\dot{\tilde{m}} &=& \left\{-\frac{\gamma_{\rm{m}}}{2}+\frac{g^{2}n_{\rm{M}}}
{-i(\Delta_{\rm{E}}+\Delta_{\rm{m}})+\frac{\gamma_{\rm{E}}}{2}}\right\}\tilde{m}.
\end{eqnarray}
From Eq. (\ref{equ:4}) we observe that the Brillouin scattering process between the optical photon and magnon in YIG sphere contributes an effective magnon gain as
\begin{eqnarray}\label{equ:5}
  \mathbb{G} = {\rm{Re}}\left\{\frac{g^{2}n_{\rm{M}}}
{-i(\Delta_{\rm{E}}+\Delta_{\rm{m}})+\frac{\gamma_{\rm{E}}}{2}}\right\} = \frac{g^{2}n_{\rm{M}}\frac{\gamma_{\rm{E}}}{2}}{(\Delta_{\rm{E}}+\Delta_{\rm{m}})^{2}+(\frac{\gamma_{\rm{E}}}{2})^{2}}.\nonumber\\
\end{eqnarray}
We define M[$\gamma_{\rm{m}}^{-1}$] = ${\rm{exp}}[(2\mathbb{G}-\gamma_{m})/\gamma_{m}]$ as the steady state number of magnons.
We can see that the magnon gain factor is
proportional to the input light power enlightening us of the possibility of implementing magnon laser by using light.
By setting $\gamma_{\rm{m}}/2 = \mathbb{G}$ , we obtain the threshold power of the magnon laser
\begin{eqnarray}\label{equ:6}
  \rm{P}_{th} &=& \frac{\hbar\omega_{\rm{M}}\gamma_{\rm{m}}\gamma_{\rm{M}}^{2}[(\Delta_{\rm{E}}+\Delta_{\rm{M}})^{2}+
  (\frac{\gamma_{\rm{E}}}{2})^{2}]}{4g^{2}\Upsilon_{\rm{M}}\gamma_{\rm{E}}},
\end{eqnarray}
referring to the incident power required when the magnon gain induced by
Brillouin scattering overcomes the dissipation of the magnon, which is the necessary condition for the generation of magnon laser.

Figure \ref{fig:2} shows the threshold power of the magnon laser
$\rm{P_{th}}$ varies with the magnon-photon coupling strength $g$ and the detuning $\Delta_{\rm{m}}/\Omega_{\rm{m}}$.
\begin{figure}[htbp]
\centering
\includegraphics [width=0.9\linewidth] {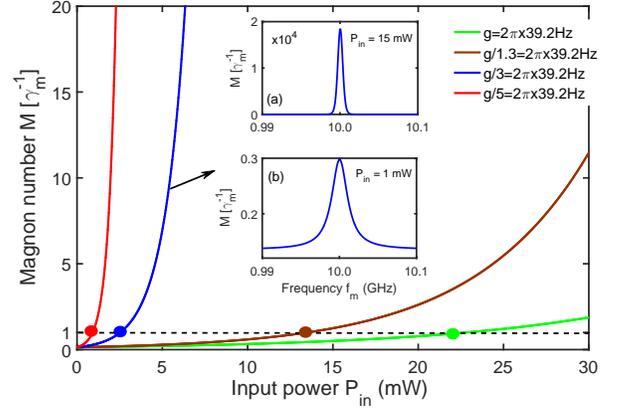}
\caption{The stimulated emitted magnon number M[$\gamma_{\rm{m}}^{-1}$] as a function of the input power in the context of different magnon-photon coupling strength $g$ under the resonance condition $\Delta_{\rm{m}}/\Omega_{\rm{m}} = 0$.
The black dotted line represents the threshold conditions for the generation of magnon laser, i.e., M[$\gamma_{\rm{m}}^{-1}$] = 1.
The threshold condition denoted by the thick points is obtained for $\gamma_{\rm{m}}/2 = \mathbb{G}$.
The insets (a) and (b) plot the magnon line shape function in the case of $\rm{P_{in}}$ = 15 and  1 $\rm{mW}$, respectively.
The other parameters are the same as those in Fig. \ref{fig:2}.}
\label{fig:4}
\end{figure}
We can clearly see that the threshold power is the lowest when the magnon mode satisfies the resonance condition, i.e., $\Delta_{\rm{m}} = \Omega_{\rm{m}} - \omega_{\rm{m}} = 0$, but will exponentially increase when it deviates from the resonance position.
For a YIG sphere with a diameter of 200-$\mu m$ \cite{strong1}, without assuming further optimization process of the YIG sphere, the magnon-photon coupling constant is theoretically evaluated to be
$g = 2\pi\times39.2$ $\rm{Hz}$ \cite{WGM1}.
In this realistic assumption, the threshold power of the magnon laser is about 22 $\rm{mW}$ (the red line shown). 
In addition, increasing the coupling strength to $g = 1.5\times2\pi\times39.2$ $\rm{Hz}$ and $g = 2.4\times2\pi\times39.2$ $\rm{Hz}$, as the yellow and green lines shown, the threshold power is correspondingly reduced from $\rm{P}_{\rm{th}} = 10$ $\rm{mW}$ to 4 $\rm{mW}$.
The physical mechanism as Eq. (\ref{equ:6}) reveals that the threshold power of the magnon laser is inversely proportional to the square of the magnon-photon coupling strength $g$.
In experiment, the improvement of the magnon-photon coupling strength might be realized in several aspects,
for instance, scaling down the YIG sphere size to reduce the mode volume \cite{WGM1},
manufacturing the nanostructured magnets \cite{device1,device2} to increase the spatial overlap between the optical and magnon fields, and further purifying and chemically processing of the YIG sphere are also recommended \cite{Brillouin2}.
With these improvements, the magnon-photon coupling strength is expected to increase two orders of magnitude \cite{EIT}, and the threshold power is decreased to
$\rm{P}_{{\rm{th}}} \sim$ 2.3 $\mu$W.
In this context, even if the input power is quite weak, the magnon laser action can be observed experimentally in the near future.

From Eqs. (\ref{equ:4}) and (\ref{equ:5}), the
stimulated emitted magnon number can be derived as M[$\gamma_{\rm{m}}^{-1}$] = ${\rm{exp}}(\eta-1)$ \cite{OMIT1}, where
\begin{figure}[htbp]
\centering
\includegraphics [width=1\linewidth] {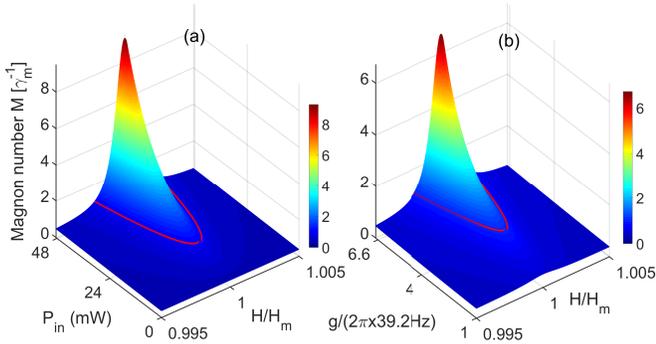}
\caption{Three-dimensional graph of the stimulated emitted magnon number M[$\gamma_{\rm{m}}^{-1}$] varies with
the magnetic field strength $\rm{H/H_{m}}$, and (a) the optical power $\rm{P_{in}}$ $(\rm{mW})$, (b) the magnon-photon coupling strength $g$.
The red solid line represents the threshold condition of the magnon laser.
We use $g = 2\pi\times39.2$ $\rm{Hz}$ in Fig. \ref{fig:4}(a) and $\rm{P_{in} = 1 mW}$ in Fig. \ref{fig:4}(b), and the
other parameters are the same as those in Fig. \ref{fig:2}.} 
\label{fig:5}
\end{figure}
\begin{eqnarray}
  \eta &=& \frac{8g^{2}\rm{P_{in}}}{\hbar\omega_{\rm{M}}\gamma_{\rm{M}}\gamma_{\rm{E}}\gamma_{\rm{m}}},
\end{eqnarray}
represents the pure gain factor of the emitted magnon number under the triple-resonance condition.
In the case of $\eta > 1$ implies that the magnon gain caused by the Brillouin light scattering overcomes the mode dissipation and yields the magnon coherent amplification.
As the green solid line shown in Fig. \ref{fig:4}, for the magnon-photon coupling strength $g = 2\pi\times39.2$ $\rm{Hz}$, the magnon laser action would be observed when the incident light power reaches about 22 $\rm{mW}$, which is highly accessible under the current experimental technique \cite{WGM1}.
More importantly, the emitted magnon number M[$\gamma_{\rm{m}}^{-1}$] increases exponentially with the incident power.
In particular, when the magnon-photon coupling strength increase to $g = 3\times2\pi\times39.2$ $\rm{Hz}$, as the blue solid line shown, the threshold power of the magnon laser is reduced to 2.5 $\rm{mW}$. And the steady state number of stimulated emitted magnons is calculated to be M[$\gamma_{\rm{m}}^{-1}$] $\approx$ $1.82\times10^{4}$ when the input light power $\rm{P}_{\rm{in}}$ = 15 $\rm{mW}$, which is much larger than the number of thermally populated magnons $m_{\rm{th}} \approx 624.8$ at room temperature (T = 300 K), consequently, the magnon laser effect can be measured at room temperature.
Besides, the magnon line shape function are plotted in the illustration (a) and (b) of Fig. \ref{fig:4} in the context of the pumping points below ($\rm{P_{in}}$ = 1 $\rm{mW}$) and above ($\rm{P_{in}}$ = 15 $\rm{mW}$) the threshold, respectively.
We can see that the stimulated emitted magnons are mainly populated in the vicinity of the frequency $\omega_{\rm{m}}$ = $2\pi \times 10$ $\rm{GHz}$ $\simeq$ 62.83 $\rm{GHz}$, and the linewidth of the magnon laser above the threshold is significantly narrower \cite{narrower}.
These results, therefore, are enlightening for the realization of magnon operation by using light, and may offer theoretical support for the manufacture of high-intensity magnon-laser amplifiers.


Finally, achieving the effective control of the magnon laser action is of fundamental importance and is also a key link for the practical application of the magnon laser in the future.
Obviously, we can see from Figure \ref{fig:5} that the stimulated magnon number reaches the peak only when the magnetic field strength was adjusted to ${\rm{H = H_{m}}} = \Omega_{\rm{m}}/\varrho \approx 357.14$ $\rm{mT}$.
In this circumstances, the frequency of the magnon mode in YIG sphere is resonant with the frequency difference between the incident photons and the scattered photons.
Otherwise, the emitted magnon number decreases exponentially when the magnetic field strength deviates from $\rm{H_{m}}$, i.e., $\rm{H}>\rm{H_{m}}$ or $\rm{H}<\rm{H_{m}}$.
Physically, on one hand, the frequency of the Kittel mode in YIG sphere is determined by the external magnetic field strength \cite{Q2}, i.e., $\Omega_{\rm{m}} = \varrho \rm{H_{m}}$.
On the other hand, as Fig. \ref{fig:2} shown, the threshold power of the magnon laser is the weakest when the optical mode as well as the magnon mode simultaneously on resonance.
The red line in Fig. \ref{fig:5} indicates the threshold conditions of the magnon laser, i.e., the gain factor $\eta = 1$.
Namely, if and only if the system parameters within the red line region, i.e., the gain factor $\eta > 1$, the magnon laser action can be achieved, in which the external magnetic field plays a paramount role.
Our proposal thus provides a pathway for implementing magnetic-field-modulated magnon laser that can be applied to both fundamental problems in cavity optomagnonics and influences a broad range of scientific fields beyond magnonics.


In conclusion, the magnon laser action based on Brillouin light scattering in a YIG ferrimagnetic sphere has been theoretically demonstrated.
We manifest that the stimulated emitted magnon number increases exponentially with the input light power, similar to the photon amplification by stimulated emission of radiation, providing a theoretical foundation for the realization of magnon laser.
The magnon laser action discussed here further demonstrates the feasibility of achieving magnon manipulation by using light.
These results, therefore, are expected to pave a path toward the achievement of microwave-to-optical converter as well as the preparation of coherent magnon sources, and may promote the rapid development of the thermodynamics and spintronics of the magnet.

\begin{acknowledgments}
This work was supported by the National Natural Science Foundation of China (NSFC) (11774113, 61775036); The high-level talent program of Dongguan University of Technology (KCYCXPT2017003).
\end{acknowledgments}

\end{document}